\documentclass[twocolumn,prl]{revtex4}
\usepackage[latin9]{inputenc}
\usepackage{color}
\usepackage{verbatim}
\usepackage{amsmath}
\usepackage{amssymb}
\usepackage{graphicx}
\usepackage{esint}
\usepackage{hyperref}

\makeatletter
\@ifundefined{textcolor}{}
{%
 \definecolor{BLACK}{gray}{0}
 \definecolor{WHITE}{gray}{1}
 \definecolor{RED}{rgb}{1,0,0}
 \definecolor{GREEN}{rgb}{0,1,0}
 \definecolor{BLUE}{rgb}{0,0,1}
 \definecolor{CYAN}{cmyk}{1,0,0,0}
 \definecolor{MAGENTA}{cmyk}{0,1,0,0}
 \definecolor{YELLOW}{cmyk}{0,0,1,0}
}

\usepackage{mathrsfs}

\draft

\makeatother

\begin{document}

\title{Non-Abelian Majorana modes protected by an emergent second Chern
number}

\author{Cheung Chan}

\affiliation{International Center for Quantum Materials, School of Physics, Peking
University, Beijing, 100871, China}

\affiliation{Collaborative Innovation Center of Quantum Matter, Beijing 100871,
China}

\author{Xiong-Jun Liu}

\thanks{Corresponding author: \href{http://xiongjunliu@pku.edu.cn}{xiongjunliu@pku.edu.cn}}

\affiliation{International Center for Quantum Materials, School of Physics, Peking
University, Beijing, 100871, China}

\affiliation{Collaborative Innovation Center of Quantum Matter, Beijing 100871,
China}
\begin{abstract}
The search for topological superconductors and non-Abelian Majorana
modes ranks among the most fascinating topics in condensed matter
physics. There now exist several fundamental superconducting phases
which host symmetry protected or chiral Majorana modes. The latter,
namely the chiral Majorana modes are protected by Chern numbers in
even dimensions. Here we propose to observe novel chiral Majorana
modes by realizing Fulde-Ferrell-Larkin-Ovchinnikov state, i.e. the
pairing density wave (PDW) phase in a Weyl semimetal which breaks
time-reversal symmetry. Without symmetry protection, the 3D gapped
PDW phase is topologically trivial. However, a vortex line generated
in such phase can host chiral Majorana modes, which are shown to be
protected by an emergent second Chern number of a synthetic 4D space
generalized from the PDW phase. We further show that these chiral
modes in the vortex rings obey 3D non-Abelian loop-braiding statistics,
which can be applied to topological quantum computation.
\end{abstract}
\maketitle
Eighty years ago, Ettore Majorana proposed a new fermion which is
identical to its antiparticle and now called Majorana fermion (MF)
\cite{Majorana2008}, and speculated that it might interpret neutrinos.
While the evidence of MFs as elementary particles in high energy physics
is yet elusive, the search for MFs has energetically revived in condensed
matter physics and become an exciting pursuit in recent years~\cite{Wilczek2009,Alicea2012RPP,Franz2013a}.
Rather than being elementary particles, MFs can emerge as quasiparticles
localized in surfaces or defects of $p$-wave topological superconductors
(SCs) \cite{PhysRevB.61.10267,Kitaev2001}. Besides the early proposed
intrinsic $p$-wave SCs, the studies have shown that combining conventional
$s$-wave SC and spin-orbit (SO) coupled systems can render effective
$p$-wave pairing states \cite{Fu2008Majorana,Sau2010Majorana,Lutchyn2010Majorana,Oreg2010Majorana,Franz2015RevModPhys},
bringing the realization of MFs to realistic solid state experiments.
In the last couple of years, the suggestive signatures of Majorana modes have been observed with
heterostructures formed by $s$-wave SCs and semiconductor nanowires
\cite{Mourik1003,Deng2012,Das2012}, magnetic chains \cite{Nadj-Perge2014},
or topological insulators \cite{Xu2015,PhysRevLett.116.257003}.

MFs in different dimensional SCs exhibit fundamentally distinct properties.
In odd dimensions topological SCs classified by integers necessitate
symmetry protection, such as time-reversal (TR) and particle-hole
symmetries, so do Majorana modes in such systems. The famous examples
include the 1D Kitaev chain \cite{1063-7869-44-10S-S29} which belongs
to BDI symmetry class, with Majorana zero modes at chain ends
being protected by chiral symmetry, and the 3D TR invariant (class
DIII) topological SC \cite{PhysRevLett.102.187001,PhysRevB.78.195125,kitaev2009,Teo2016RevModPhys}
which hosts helical Majorana surface modes protected by TR symmetry
and characterized by a 3D winding number. In contrast, topological
SCs in even dimensions can be chiral topological orders which require no symmetry protection. A notable toy model is the 2D $p+ip$
SC \cite{PhysRevB.61.10267,PhysRevLett.86.268} which is a non-Abelian
chiral topological order classified by the 1st Chern number, and hosts
chiral Majorana edge modes whose gapping is forbidden by energy conservation.
Stacking $p+ip$ SCs along the third direction yields a 3D system~\cite{PhysRevB.82.115120}
with anisotropic Majorana surface modes corresponding to, however,
the Chern numbers of the original 2D subspace rather than any 3D topological
invariant, implying that these surface modes are not intrinsic chiral
states of the 3D phase. As a consequence, the stacked phase is not
a 3D non-Abelian topological order, unlike the 2D $p+ip$ chiral phases.

Here we propose to observe novel chiral Majorana modes by
realizing Fulde-Ferrell-Larkin-Ovchinnikov state \cite{Fulde1964,Larkin1965},
i.e. pairing density wave (PDW) phase in a 3D Weyl semimetal whose
study attracted great interests only recently \cite{PhysRevB.83.205101,PhysRevLett.107.127205,Xu613,Huang2015,Lv2015,PhysRevX.5.031013,PhysRevLett.114.237001},
and show that such modes can realize exotic non-Abelian loop braiding
statistics, applicable to topological quantum computation \cite{Kitaev20032,RevModPhys.80.1083,Alicea2011Nat}.
While the 3D gapped PDW phase is topologically trivial, a vortex line
generated in this phase can host intrinsic chiral MFs which are protected by an emergent 2nd Chern number of
a synthetic 4D space generalized from the physical 3D system. This
study shows that the gapped PDW phase, being trivial in 3D, is
an emergent non-Abelian topological order in a 4D synthetic space.

\begin{figure*}
\includegraphics[width=0.85\textwidth]{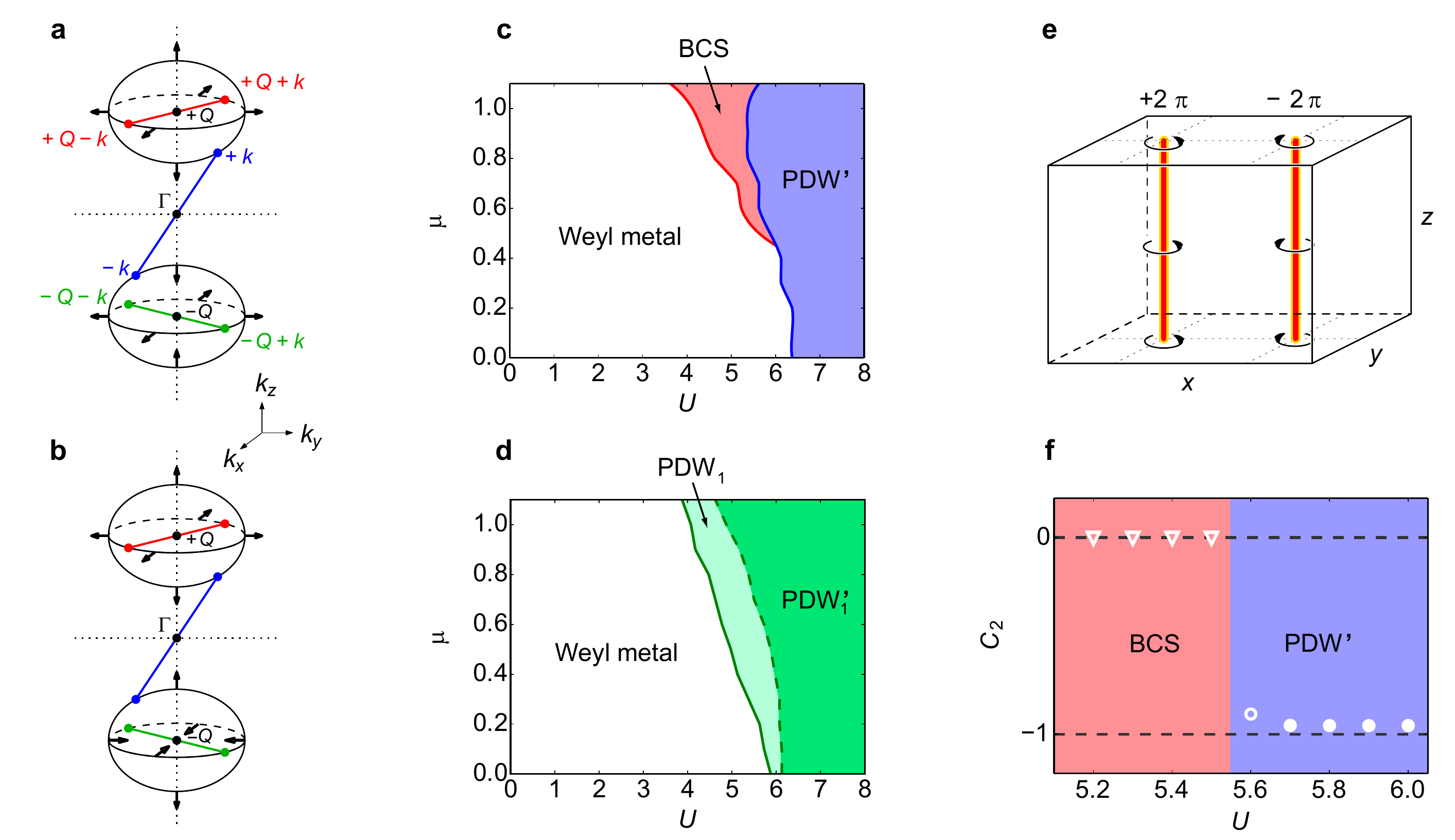}

\caption{\label{fig:Weyl}Weyl metals, mean field phase diagrams, vortex line
configuration, and 2nd Chern numbers. (a,b) Schematic for Fermi surfaces
and superconducting pairing of inversion symmetric (for a) and inversion
symmetry broken (for b) Weyl metals. The two ellipsoids centered at
$\mathbf{Q}_{\pm}$ are the Fermi surfaces. The spin orientations,
shown by black arrows, at the Fermi surface around $\mathbf{Q}_{-}$
is flipped when $0<\mu<2t_{z}^{s}\sin|Q_{\pm}|$. The colored lines
represent the pairing order parameters $\Delta_{\mathbf{q}}$ with
$\mathbf{q}=0,2\mathbf{Q}_{\pm}$. (c,d) Phase diagram for inversion
symmetric (for c) and inversion symmetry broken (for d) Weyl metal
versus attractive Hubbard interaction $U$ and chemical potential
$\mu$. (e) Configuration of two vortex lines with opposite vorticities.
(f) The emergent 2nd Chern number $C_{2}$ for a 4D synthetic space
generalized from the 3D physical system in the inversion symmetric
case and with $\mu=0.7$. Upon increasing $U$, the system undergoes
a transition from BCS to $\mathrm{PDW}^{\prime}$ phase. The circles
(triangles) gives the 2nd Chern number in BCS ($\mathrm{PDW}^{\prime}$)
phase, which shows $C_{2}\approx-0.956$ for $\mathrm{PDW}^{\prime}$
phase.}
\end{figure*}

\textit{Model.\textendash }We start with a minimal model for Weyl
semimetal tuned by SO coupling, which can be realized for cold atoms
based on a recent experiment~\cite{Wang2016,Wu2016PKU}, together
with an attractive Hubbard interaction of strength $U$ which can drive superconducting/superfluid
phases. The present Weyl-Hubbard model is described by
\begin{eqnarray}
H & = & \sum_{\mathbf{p}}\psi_{\mathbf{p}}^{\dagger}h_{\mathbf{p}}\psi_{\mathbf{p}}-U\sum_{\mathbf{i}}n_{\mathbf{i}\uparrow}n_{\mathbf{i}\downarrow},\label{eq:ham_0}\\
h_{\mathbf{p}} & = & [m_{z}-2t_{0}(\cos p_{x}+\cos p_{y})-2t_{z}^{c}\cos p_{z}]\sigma_{z}\nonumber \\
 &  & +2t_{{\rm SO}}(\sin p_{x}\sigma_{x}+\sin p_{y}\sigma_{y})-\mu-2t_{z}^{s}\sin p_{z},\nonumber
\end{eqnarray}
where the spinor $\psi_{\mathbf{p}}=(c_{\mathbf{p}\uparrow},c_{\mathbf{p}\downarrow})$, the particle number operator
$n_{\mathbf{i}\sigma}=c_{\mathbf{i}\sigma}^{\dagger}c_{\mathbf{i}\sigma}$, $t_{0}$ ($t_{{\rm SO}}$) is the
coefficient of spin-conserved (spin-flipped) hopping in $x$-$y$
plane, $\sigma_{x,y,z}$ are Pauli matrices on spin space, and $\mu$
is chemical potential.
Moreover, we consider the hopping terms along $z$ direction with
$(t_{z}^{c},t_{z}^{s})=t_{z}(\cos\varphi_{0},\sin\varphi_{0})$,
which break inversion symmetry of the Weyl semimetal unless $\varphi_{0}=n\pi/2$
with integer $n$. For $m_{z}=4t_{0}+2t_{z}^{c}\cos Q$ ($0<Q<\pi$),
the Weyl semimetal has two nodal points of chiralities $\chi=\pm$
located at $\mathbf{Q}_{\chi}=(0,0,\chi Q)$ and with energies $E_{\pm}=-\mu\mp2t_{z}^{s}\sin Q$,
respectively. Without loss of generality, in the following we shall
choose $Q=\frac{2\pi}{3}$ and $t_{0,z}=t_{{\rm SO}}=1.0$ to facilitate
further discussion.

The superconducting/superfluid phases can be induced with an attractive interaction $U>0$. The possible
pairing orders are of two distinct types, namely the $s$-wave BCS
and PDW phases, as sketched in Fig.~\ref{fig:Weyl}(a). The former
describes a uniform pairing order $\Delta_{0}$ occurring between
two different Weyl cones and with zero center-of-mass momentum of Cooper
pairs, while the latter are spatially modulated
orders $\Delta_{2\mathbf{Q}_{\pm}}$ occurring within each Weyl cone
and the Cooper pairs have nonzero center-of-mass momentum~\cite{PhysRevLett.114.237001,PhysRevB.86.214514,PhysRevB.92.035153,PhysRevB.93.094517,PhysRevB.94.075115}.
These pairing orders can be written as $\Delta_{\mathbf{q}}=\frac{U}{2N}\sum_{\mathbf{k}}\langle c_{\mathbf{q}/2+\mathbf{k}\uparrow}c_{\mathbf{q}/2-\mathbf{k}\downarrow}\rangle$,
with $\mathbf{q}=0,2\mathbf{Q}_{\pm}$, and $N$ being number
of sites, and the interaction is decoupled into $\mathcal{H}_{\mathrm{MF}}=\sum_{\mathbf{k},\mathbf{q}}\Delta_{\mathbf{q}}\thinspace c_{\mathbf{q}/2+\mathbf{k}\uparrow}^{\dagger}c_{\mathbf{q}/2-\mathbf{k}\downarrow}^{\dagger}+\mathrm{h.c}$.
Here $\mathbf{k}$ is summed over the entire Brillouin
zone. Having both BCS and PDW phases, the order parameter takes
the following generic form in real space
\begin{equation}
\Delta(\mathbf{r})=\Delta_{0}+\Delta_{+2Q}e^{i2\mathbf{Q}_{+}\cdot\mathbf{r}}+\Delta_{-2Q}e^{i2\mathbf{Q}_{-}\cdot\mathbf{r}}.\label{eq:Delta_x}
\end{equation}
The BCS
and PDW orders may compete governed by following underlying mechanisms~\cite{SI}. The BCS pairing connects two different Weyl cones with opposite
chiralities and cannot fully gap out the bulk, since the chiralities,
characterizing the topology of the Weyl Fermi surfaces, cannot be
adiabatically connected without closing the bulk gap. Accordingly,
the BCS phase renders a Weyl SC with four nodal points \cite{PhysRevB.86.214514}.
In contrast, the PDW order can fully gap out the bulk since the pairings
do not connect the two Weyl cones but fold up the original Brillouin
zone due to its spatial modulation. Nevertheless, such spatial modulation
may cost extra energy. Below we perform a self-consistent study of
the complete phase diagram in different regimes.

The numerical simulation reveals different phase diagrams versus $\mu$
and $U$ for the inversion-symmetric {[}Fig.~\ref{fig:Weyl}(b){]}
and inversion symmetry broken {[}Fig.~\ref{fig:Weyl}(d){]} Weyl
metals. For the inversion-symmetric Weyl metal given by $\varphi_{0}=0$,
a direct transition from the Weyl metal phase to the $\mathrm{PDW}'$
phase, which has $|\Delta_{+2Q}|=|\Delta_{-2Q}|\gg|\Delta_{0}|\neq0$,
is obtained by increasing $U$ in the low chemical potential regime
with $|\mu|<\mu_{c}$ and $\mu_{c}\sim0.4$. The equality $|\Delta_{+2Q}|=|\Delta_{-2Q}|$
in the $\mathrm{PDW}'$ phase is a consequence of the inversion symmetry.
However, when $\mu$ is tuned away from Weyl point and beyond the
critical value $|\mu|>\mu_{c}$, the BCS phase with $\Delta_{0}\neq0$
and $\Delta_{\pm2Q}=0$ appears between the Weyl metal and $\mathrm{PDW}'$
phase. This result reflects that the pairings within each Weyl Fermi
surface dominates over those between
two different Weyl pockets in relatively low chemical potential regime.
On the other hand, if the inversion symmetry is broken, the BCS
phase is suppressed, and the system enters from Weyl metal into
$\mathrm{PDW}_{1}$ state first and then $\mathrm{PDW}_{1}^{\prime}$
phase by increasing $U$, as shown in Fig.~\ref{fig:Weyl}(d) with
$\varphi_{0}=\frac{3}{32}\pi$. The $\mathrm{PDW}_{1}$ phase is characterized
by $|\Delta_{+2Q}|\geq|\Delta_{-2Q}|$ for $\mu\geq0$ and $\Delta_{0}=0$,
while in the $\mathrm{PDW}_{1}^{\prime}$ phase a small BCS order
$|\Delta_{0}|<|\Delta_{\pm2Q}|$ also emerges.

\begin{figure*}
\includegraphics[width=0.99\textwidth]{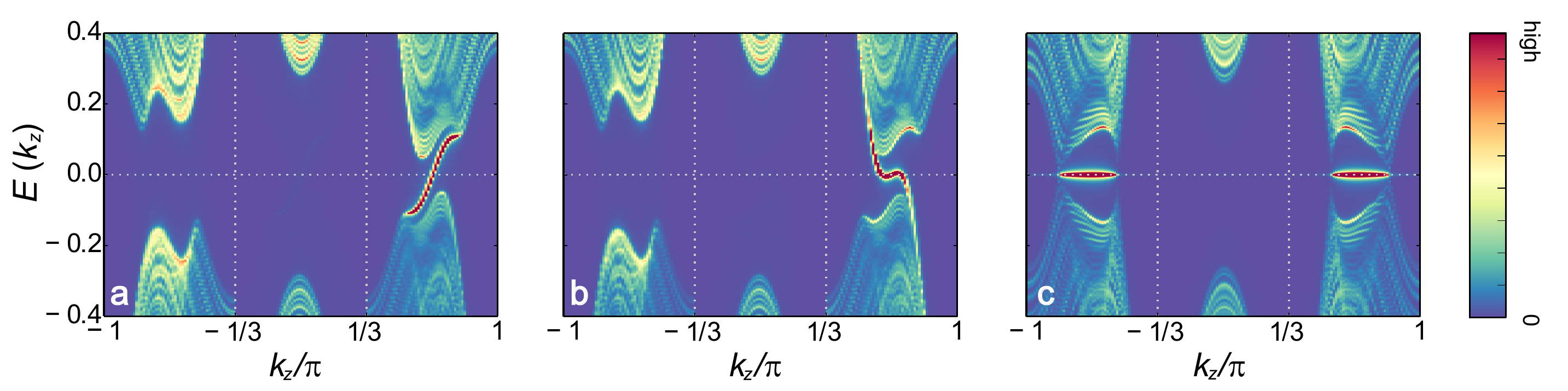}

\caption{\label{fig:lineModes}Chiral Majorana modes as in-gap bound states
to a vortex line. (a-b) Spectral functions $A(k_{z},E)$ in the inversion
symmetric Weyl metal with PDW orders obtained from self consistent
mean field for $\mu=0.7$ and $U=5.8$. The vortex line is attached
to $\Delta_{+2Q}$, with winding number $n=+1$ for (a) and $n=-1$
for (b). (c) Spectral function $A(k_{z},E)$ for the BCS dominated
phase in the inversion symmetric Weyl metal with $\Delta_{0}=0.25$.
Two segments of Majorana flat bands are obtained.}
\end{figure*}

\textit{Vortex Line Modes.\textendash }Both BCS and PDW states
have only the particle-hole symmetry, so they are
class D SCs according to the Altland-Zirnbauer symmetry classification~\cite{Altland-Zirnbauer}.
From the current topological classification theory \cite{PhysRevB.78.195125,kitaev2009,Teo2016RevModPhys},
both BCS and PDW phases are $p_{x}+ip_{y}$ SCs stacked
along $z$ direction. Nevertheless, the two
phases exhibit fundamental distinction in the excitations along vortex lines generated to the pairing orders.

We consider first the vortex line excitations for the PDW dominated
phase. For simplicity, we take that $|\Delta_{+2Q}|=|\Delta_{-2Q}|$
and $\Delta_{0}=0$, since a small perturbative BCS order does not
affect the results. Let a vortex line with winding number $n$ be
along $z$-axis and attached to $\Delta_{+2Q}$, so that
$\Delta_{+2Q}=|\Delta_{+2Q}|\exp[in\phi(\mathbf{r})]$, where the
azimuthal angle $\tan(\phi)=y/x$. In the low energy limit we can
write down the effective Hamiltonian by linearizing the 3D Weyl cone
Hamiltonian around $\mathbf{Q}_{+}$ point
\begin{align}
H_{\mathrm{eff}}= & -(\sum_{j=x,y,z}iv_{j}\sigma_{j}{\partial}_{j}+\mu_{\mathrm{eff}})\tau_{z}\label{eq:H_eff}\\
 & +\bar{\Delta}(\rho)\cos(n\phi)\tau_{x}+\bar{\Delta}(\rho)\sin(n\phi)\tau_{y},\nonumber
\end{align}
where $v_{j}$ is the Fermi velocity along $j$-th direction, $\mu_{\mathrm{eff}}$
is the chemical potential measured from the Weyl node, and $\tau$'s
are Pauli matrices for the Nambu space spanned by $\hat{f}(\mathbf{r})=[c_{\uparrow}(\mathbf{r}),c_{\downarrow}(\mathbf{r}),c_{\downarrow}^{\dagger}(\mathbf{r}),-c_{\uparrow}^{\dagger}(\mathbf{r})]^{T}$.
We implicitly assumed that the vortex line is located at the origin
in the $(x,y)$-plane along $z$-axis, hence we choose cylindrical
coordinate $(\rho,\phi,z)$. Here $\bar{\Delta}$ is a function that
$\bar{\Delta}(\rho\rightarrow0)=0$ and $\bar{\Delta}(\rho\rightarrow\infty)=\mathrm{constant}$
and $n\phi$ is the phase winding of the PDW component. As detailed
in the Supplementary Material~\cite{SI}, for $n=\pm1$ the Majorana in-gap
modes can be obtained analytically and satisfy $H_{\mathrm{eff}}\gamma_{z}(\rho,\phi,k_{z})={\cal E}_{k_{z}}\gamma_{z}(\rho,\phi,k_{z})$,
with eigenvalues ${\cal E}_{k_{z}}={\rm sgn}(n){\rm sgn}(v_{x}v_{y})v_{z}k_{z}$,
which implies that the vortex Majorana modes are chiral. The Majorana
operator can be constructed by $\hat{\gamma}_{z}(k_{z})=\int\!d^{2}\mathbf{r}\thinspace\gamma_{z}(\rho,\phi,k_{z})\hat{f}(\rho,\phi,k_{z})$,
with $\hat{\gamma}_{z}(k_{z})=\hat{\gamma}_{z}^{\dagger}(-k_{z})$
for real MFs. It can be seen that the chirality $\chi_{\mathrm{M}\ell}={\rm sgn}(nv_{x}v_{y}v_{z})$
of the vortex Majorana modes related to the vortex winding $n$ and
Weyl node chirality $\chi$. With the chiral properties of Majorana
modes the above solution can be generalized to the case with a generic
winding number $n={\cal N}$. Actually, such a vortex line is topologically
equivalent to $|{\cal N}|$ vortex lines with unity winding $n={\rm sgn}({\cal N})$.
In the later case each vortex line hosts a branch of chiral Majorana
modes. Due to the chiral property putting the $n$ branches of vortex
modes together cannot annihilate them, yielding $n$ chiral Majorana
vortex modes. Nevertheless, in the continuous limit a Majorana zero
mode is obtained at $k_{z}=0$ only when $n$ is an odd integer.

The chirality of Majorana vortex modes imply that these modes are
gapless and traverse the bulk gap of the PDW phase. To confirm this
result, we perform a full real-space numerical calculation of the
Majorana modes based on the lattice model. In particular, we consider
the inversion symmetric Weyl metal with $\mu=0.7$ and $U=5.8$. The
self-consistent calculation reveals a $\mathrm{PDW}^{\prime}$ phase
with $\Delta_{\pm2Q}\approx0.201$ and $\Delta_{0}\approx-1.85\times10^{-2}$,
and the system has a bulk gap $E_{{\rm gap}}\approx0.21$. We consider
a vortex line ($n=1$) and anti-vortex line ($n=-1$) along $z$-axis,
separated from each other in $x$-$y$ plane, and attached to one
of $\Delta_{0,\pm2Q}$, as sketched in Fig.~\ref{fig:Weyl}(e). With
this configuration appropriate periodic boundary condition can be
applied in the numerical calculation. The local spectral function
$A(x,y,k_{z},E)$ can be obtained from the retarded Green's function
$G^{R}(E)$ of the system
\begin{equation}
A=-\frac{1}{\pi}\sum_{s=\uparrow,\downarrow}{\Im}\langle x,y,k_{z},s|G^{R}(E)|x,y,k_{z},s\rangle,
\end{equation}
where $|x,y,k_{z},s\rangle$ is the Bloch basis with momentum $k_{z}$
and in real space for $x$-$y$ plane~\cite{SI}. Computing $A(k_{z},E)$ near
the vortex core gives the energy spectra of the bulk and the corresponding
vortex lines.

Fig.~\ref{fig:lineModes}(a) and (b) show the spectra measured from
the vortex line ($n=1$) and anti-vortex line ($n=-1$), respectively.
It is clear that in both cases the chiral Majorana vortex modes traverse
the bulk gap connecting the lower and upper bands. The chirality
of these modes depends on the vortex line winding number,
consistent with the previous analytic solution. This result is fundamentally
different from that for a BCS dominated phase, as shown in Fig.~\ref{fig:lineModes}(c),
where we compute the Majorana modes by attaching vortex and anti-vortex
lines to $\Delta_{0}$ with $\Delta_{0}=0.25$ and $\Delta_{\pm2Q}=0$.
Majorana zero-energy flat bands associated with the vortex lines are
obtained. These Majorana zero modes are simply the vortex modes of
$p_{x}+ip_{y}$ SCs with different momenta $k_{z}$.

\textit{Emergent 2nd Chern number.\textendash }Chiral gapless modes have to be protected by chiral topological invariants,
e.g. the Chern numbers. However, it can be verified that for any
2D sub-plane incorporating $z$-axis the 1st Chern number of the
PDW phase is zero. Moreover, the 3D system is also topologically
trivial \cite{PhysRevB.78.195125,kitaev2009} without symmetry
protection. As a result, a higher Chern number~\cite{PhysRevB.78.195424,PhysRevB.82.115120,Bernevig2013} may protect
the Majorana vortex line modes. Note that the SC order can be parameterized
by its phase factor $\Delta_{\mathbf{q}}e^{in\theta}$, where $\theta\in[0,2\pi)$
forms a 1D periodic parameter space $S^{1}$. Together with the 3D
lattice, we construct a 4D synthetic space $T^{4}=T^{3}\times S^{1}$
spanned by $\mathsf{p}=(\mathbf{p},p_{\theta})$ with $\mathbf{p}=(p_{x},p_{y},p_{z})$
and $p_{\theta}=\theta$. In this synthetic 4D space we define the
2nd Chern number by
\begin{equation}
C_{2}=\frac{1}{32\pi^{2}}\int_{T^{4}}d^{4}\mathsf{p}\thinspace\epsilon_{ijk\ell}\mathrm{Tr}[\mathtt{F}_{ij}\mathtt{F}_{k\ell}]\in\mathbb{Z},
\end{equation}
where $\epsilon$ is the antisymmetric tensor and $\mathtt{F}_{ij}$
are the gauge field strengths calculated from the eigenvectors of
the parameterized mean-field Hamiltonian $H(\mathbf{p},p_{\theta})$.
We calculate the 2nd Chern number for BCS dominated and PDW
phases, as shown numerically in Fig.~\ref{fig:Weyl}(f). It is found
that $C_{2}=0$ for the BCS dominated phase, while $C_{2}\approx-0.956$
for the PDW phase obtained with the same parameters except
for $U$ as in Fig.~\ref{fig:lineModes}(a). The deviation of $C_{2}$
from an integer is due to finite size effect~\cite{SI}.

The difference in the 2nd Chern number $C_{2}$ uncovers the essential
distinction between the BCS and PDW phases realized in a Weyl semimetal.
The BCS phase is a reminiscent of 3D quantum Hall effect which hosts
chiral edge states and dislocation line modes, while these modes are
all essentially protected by 1st Chern number, rather than 2nd
Chern number, in 2D sub-planes~\cite{PhysRevB.82.115120}. In contrast,
the protection of chiral Majorana vortex modes by the emergent 2nd
Chern number shows that the gapped PDW phase, while being trivial
in the 3D, \textit{is an intrinsic topological order in the 4D synthetic
space.}

\begin{figure}[h]
\includegraphics[width=0.95\columnwidth]{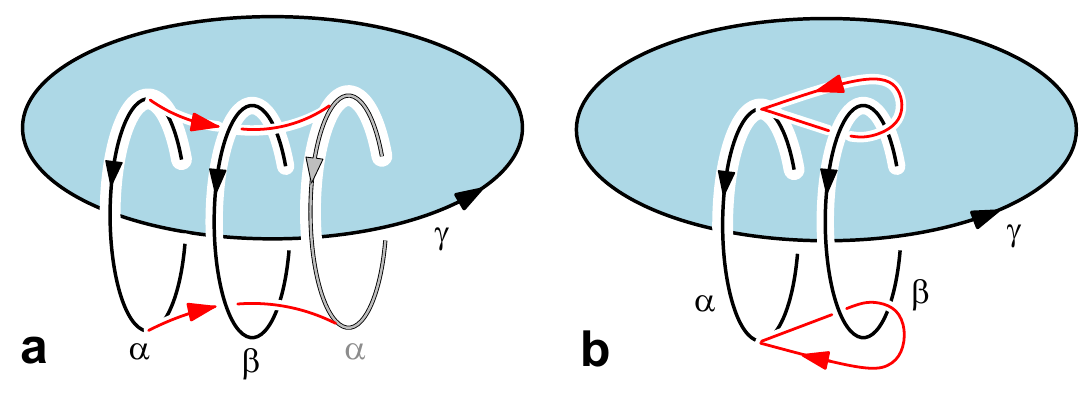}
\caption{\label{fig:3loop}Three-loop braiding of chiral Majorana modes. (a)
Self braiding of the loops $\alpha$ and $\beta$, pierced by the
third loop~$\gamma$. (b) Full braiding of loops $\alpha$ and $\beta$.}
\end{figure}

\textit{Non-Abelian loop braiding statistics.\textendash }Interpreting
the gapped PDW phase as an intrinsic topological order in the 4D synthetic
space is of high nontriviality. An important consequence is that the
chiral Majorana modes may obey new type non-Abelian statistics, for
which we consider vortex rings with winding number $n=1$ and perform
a three-loop braiding \cite{PhysRevLett.113.080403} as sketched in
Fig.~\ref{fig:3loop}. From the particle-hole symmetry and chirality
of the Majorana modes, one can readily show that the Majorana spectrum of
a vortex ring, which is threaded by even or odd number of vortex lines,
is generically given by $E_{l}=(l+1/2){\cal E}_{0}(L)$ or $E_{l}=l{\cal E}_{0}(L)$,
where $l$ is integer and ${\cal E}_{0}(L)$ depends on the vortex
ring size $L$. A vortex ring in the former case contains even number
of chiral Majorana modes without zero energy state, while in the later
case each vortex ring harbors odd number of modes including a Majorana
zero mode, which is essential for the realization of non-Abelian loop
braiding statistics.

From the above result we know that the vortex rings $\alpha$ and
$\beta$ in Fig.~\ref{fig:3loop}, which are linked to $\gamma$
separately, host one Majorana zero mode in each ring, while the vortex
ring $\gamma$, which has an even linking number, harbors no zero
mode. In this case, let $\psi_{j}$ ($j=\alpha,\beta,\gamma$) denote
the $j$-th vortex ring state which includes all chiral Majorana modes.
Under a full three-loop braiding between $\alpha$ and $\beta$, pierced
by $\gamma$ {[}Fig.~\ref{fig:3loop}(b){]}, we can show that $\psi_{\alpha},\psi_{\beta}\rightarrow-\psi_{\alpha},-\psi_{\beta}$,
while $\psi_{\gamma}$ is unchanged. A single braiding is then followed
by $\psi_{\alpha}\rightarrow\psi_{\beta}$ and $\psi_{\beta}\rightarrow-\psi_{\alpha}$
{[}Fig.~\ref{fig:3loop}(a){]}, which shows the braiding operator
to be ${\cal B}_{\alpha\beta}(\gamma)=\exp(\frac{\pi}{4}\psi_{\alpha}\psi_{\beta})$~\cite{PhysRevLett.86.268},
giving 3D non-Abelian loop braiding statistics. We note that taking
different vortex ring configurations may bring rich different non-Abelian
loop braiding statistics.

In conclusion, we have uncovered an emergent 4D non-Abelian topological
order based on a superconducting Weyl semimetal, which hosts chiral
Majorana modes protected by the 2nd Chern number. This topological
phase is distinct from the known topological superconductors in the
1D to 3D systems, and is beyond the ten-fold Altland-Zirnbauer symmetry
classification (Discussion in~\cite{SI}). Thus our results may show insight into the search
for new non-Abelian topological states. Moreover, the present result
reveals a real physical system to explore the exotic 3D non-Abelian
loop-braiding statistics, and shall push forward the studies in both
theory and experiment based on realistic solid state materials and
cold atom platforms.

We thank Frank Wilczek and Z. D. Wang for valuable discussions. This work is supported
by MOST (Grant No. 2016YFA0301604), NSFC (No. 11574008), and Thousand-Young-Talent
Program of China.

\onecolumngrid

\renewcommand{\thesection}{S-\arabic{section}}
\setcounter{section}{0}  
\renewcommand{\theequation}{S\arabic{equation}}
\setcounter{equation}{0}  
\renewcommand{\thefigure}{S\arabic{figure}}
\setcounter{figure}{0}  

\indent

\section*{Supplemental Material\\
Non-Abelian Majorana modes protected by an emergent second Chern number}

\section*{I. Self-consistent mean field study}

\section*{A. Brillouin zone folding}

The Weyl-Hubbard model can be solved self-consistently by introducing
both the BCS and PDW pairing orders. The mean field Hamiltonian is
given by
\begin{eqnarray}
H_{\mathrm{MF}} & = & \sum_{\mathbf{p}}\psi_{\mathbf{p}}^{\dagger}h_{\mathbf{p}}\psi_{\mathbf{p}}+\mathcal{H}_{\mathrm{MF}}\label{eq:H_MF}\\
h_{\mathbf{p}} & = & [m_{z}-2t_{0}(\cos p_{x}+\cos p_{y})-2t_{z}^{c}\cos p_{z}]\sigma_{z}\nonumber \\
 &  & +2t_{{\rm SO}}(\sin p_{x}\sigma_{x}+\sin p_{y}\sigma_{y})-\mu-2t_{z}^{s}\sin p_{z}\label{eq:H_MF2}\\
\mathcal{H}_{\mathrm{MF}} & = & \sum_{\mathbf{k},\mathbf{q}}\Delta_{\mathbf{q}}\thinspace c_{\mathbf{q}/2+\mathbf{k}\uparrow}^{\dagger}c_{\mathbf{q}/2-\mathbf{k}\downarrow}^{\dagger}+\mathrm{h.c.}\label{eq:H_MF3}\\
\Delta_{\mathbf{q}} & = & \frac{U}{2N}\sum_{\mathbf{k}}\langle c_{\mathbf{q}/2+\mathbf{k}\uparrow}c_{\mathbf{q}/2-\mathbf{k}\downarrow}\rangle,\label{eq:H_MF4}
\end{eqnarray}
where the Hubbard interaction is attractive for $U>0$, and the parameter
$\varphi_{0}$ governs the inversion symmetry of the system, with
$t_{z}^{s}=t_{z}\sin\varphi_{0}$, $t_{z}^{c}=t_{z}\cos\varphi_{0}$,
$m_{z}=4t_{0}+2t_{z}^{c}\cos Q$, and the spinor operator $\psi_{\mathbf{p}}=(c_{\mathbf{p}\uparrow},c_{\mathbf{p}\downarrow})$.
The BCS order corresponds to $\mathbf{q}=0$ and PDW order corresponds
to $\mathbf{q}=2\mathbf{Q}_{\pm}$, with $\mathbf{Q}_{\pm}=\pm Q\hat{e}_{z}$.
The PDW order breaks the translational symmetry by connecting two
Bloch states (one particle and one hole) with momenta $\mathbf{k}$
and $-\mathbf{k}+2Q\hat{e}_{z}$, respectively. An importance consequence
is that the original Brillouin zone (BZ) folds up into $2\pi/Q$ sub-Brillouin
zones if $Q$ is a commensurate momentum.

For convenience we choose here $Q=2\pi/3$ and $t_{0,z}=t_{{\rm SO}}=1$
for the present study. The different choice of $Q$ will not qualitatively
change the present results. With the PDW order the reduced BZ is $1/3$
of the original BZ. The Hamiltonian can then be written in the form
\begin{eqnarray}
H_{\mathrm{MF}}=\frac{1}{2}\sum_{k}\tilde{\psi}_{k}^{\dagger}\mathcal{H}_{Q,k}\tilde{\psi}_{k},\ \ \mathcal{H}_{Q,k}=\left[\begin{array}{cc}
h_{Q,k} & \hat{\Delta}\\
\hat{\Delta}^{\dagger} & -h_{Q,-k}^{T},
\end{array}\right]
\end{eqnarray}
where we take the reduced BZ as $k_{x,y}\in[-\pi,\pi)$ and $k_{z}\in[-\pi/3,\pi/3)$,
and denote basis for the reduced BZ by
\[
\tilde{\psi}_{k}^{T}=\left(c_{Q+k\uparrow},c_{k\uparrow},c_{-Q+k\uparrow},(\uparrow\rightarrow\downarrow);c_{Q-k\uparrow}^{\dagger},c_{-k\uparrow}^{\dagger},c_{-Q-k\uparrow}^{\dagger},(\uparrow\rightarrow\downarrow)\right).
\]
The explicit form of $h_{Q,k}$ is obtained by restricting the momentum
of the Bloch Hamiltonian $h_{\mathbf{p}}$ within a sub-BZ. The order
parameter $\hat{\Delta}$ in the matrix form reads
\begin{eqnarray}
\hat{\Delta}=\left[\begin{array}{cc}
 & \Delta_{[Q]}\\
-\Delta_{[Q]}
\end{array}\right],\ \ \Delta_{[Q]}=\left[\begin{array}{ccc}
\Delta_{+2Q} & \Delta_{-2Q} & \Delta_{0}\\
\Delta_{-2Q} & \Delta_{0} & \Delta_{+2Q}\\
\Delta_{0} & \Delta_{+2Q} & \Delta_{-2Q}
\end{array}\right].
\end{eqnarray}
We note that this system with reduced translational symmetry obeys
a particle-hole symmetry defined as
\[
\tau_{x}\mathcal{H}_{Q,k}\tau_{x}=-\mathcal{H}_{Q,-k}^{T},
\]
implying that the states at $(k,E)$ and $(-k,-E)$ in the reduced
BZ, instead of the unfolded BZ, form a particle-hole pair. Here $\tau_{x}$
is the Pauli matrix acting on the Nambu space.

\section*{B. Mean field phase diagram}

Utilizing Eqs.\ \eqref{eq:H_MF} to \eqref{eq:H_MF4}, we can iteratively
compute the eigenvectors of the Hamiltonian and hence $\Delta_{\mathbf{q}}$'s
until convergence. The resulting mean field phase diagrams against
attractive interaction strength $U$ and chemical potential $\mu$
are shown in Figs.\ 1(c) and (d) in the main text for the inversion
symmetric ($\varphi_{0}=0$) and inversion symmetry broken ($\varphi_{0}=\frac{3}{32}\pi$)
Weyl metals respectively. Since the phase diagrams are symmetric with
respect to $\mu\rightarrow-\mu$ due to a ``particle-hole symmetry''
that the Hamiltonian \eqref{eq:H_MF} possesses, so only the results
for $\mu\geq0$ is presented. We also restrict ourselves to $\mu<1$
such that the systems resemble a Weyl metal and hence the choice $\varphi_{0}=\frac{3}{32}\pi$
for the inversion symmetry broken Weyl metal. For $\mu\gtrsim1$,
the systems are always gapless regardless of the presence of BCS or
PDW orders and hence no well-defined gapped topological state for
defect modes.

The mean field phase diagrams show that the presence of superconducting
orders require a finite $U$. For small $\mu$, this can be
understood in terms of the vanishing density of states near the Fermi
energy. Moreover, the non-ideal nesting of Fermi surfaces is another
reason for the presence of orders only with finite $U$ \cite{PhysRevB.93.094517}.
This is in sharp contrast with the mean field analysis using Weyl
nodes with ideal nesting condition \cite{PhysRevB.86.214514}.
One can indeed improve the superconducting nesting by choosing, e.g.,
$Q=\pi/2$ (actually ideal nesting in this case). However, in the
case of $Q=\pi/2$, either one of $\Delta_{\pm2Q}$ can already gap
out the whole system. Attaching vortex line to one PDW order would
induce two chiral Majorana modes of opposite chiralities separately
at two Weyl nodes, and hence a null second Chern number. This is undesirable
for the study of the chiral Majorana modes that are nonetheless the
main focus of the paper.

\section*{II. Analytic results of Majorana modes in the low energy limit}

In this section, we analytically derive the vortex line chiral Majorana
modes in a low energy effective Hamiltonian for a single node in the
Weyl metal with pairing of one PDW component. Consider the effective
Hamiltonian
\begin{equation}
H_{\mathrm{eff}}=-(\sum_{j=x,y,z}iv_{j}\sigma_{j}\partial_{j}+\mu_{\mathrm{eff}})\tau_{z}+\bar{\Delta}(r)\cos(n\phi)\tau_{x}+\bar{\Delta}(r)\sin(n\phi)\tau_{y},\label{eq:H_eff}
\end{equation}
where $\mu_{\mathrm{eff}}$ is the effective chemical potential away
from the Weyl node and $\tau$'s ($\sigma$'s) are the Pauli matrices
for the Nambu (spin) space spanned by $f^{T}(\mathbf{r})=[c_{\uparrow}(\mathbf{r}),c_{\downarrow}(\mathbf{r}),c_{\downarrow}^{\dagger}(\mathbf{r}),-c_{\uparrow}^{\dagger}(\mathbf{r})]$.
Note that the anisotropy of Fermi velocities $v_{x,y,z}$ cannot affect
qualitatively the Majorana modes bounded to vortex lines, we can simply
take that $|v_{x}|=|v_{y}|=|v_{z}|=v_{f}$ to facilitate the discussion~\cite{Liu2014PRB}.
For this we further absorb the Fermi velocities into the spatial derivative
operator by taking $-iv_{j}\sigma_{j}\partial_{j}\rightarrow-i\sigma_{j}\partial_{j}$,
which can be easily restored later. The vortex line is located at
the origin in the $(x,y)$-plane along the $z$-axis, hence the choice
of cylindrical coordinate $(\rho,\phi,z)$. Here $\bar{\Delta}$ is
a function that $\bar{\Delta}(\rho\rightarrow0)=0$ and $\bar{\Delta}(\rho\rightarrow\infty)=\mathrm{constant}$
and $n\phi$ is the phase winding of the PDW component. $H_{\mathrm{eff}}$
can be solved in several steps. First, we perform a spin rotation
$U=\exp(-i\frac{\pi}{4}\sigma_{z})$ such that
\begin{eqnarray}
U^{-1}\sum_{j}\sigma_{j}\partial_{j}U=\partial_{x}\sigma_{y}-\partial_{y}\sigma_{x}+\partial_{z}\sigma_{z}.
\end{eqnarray}
Secondly, we solve $U^{-1}H_{\mathrm{eff}}U$ for a special case $k_{z}=0$
(i.e.\ ignore the $-i\partial_{z}\sigma_{z}\tau_{z}$ term). The
details can be found in \cite{Liu2014PRB}. The Majorana wave function
is $U^{-1}\langle\rho,\phi|\gamma_{0}\rangle$. We then perform the
inverse spin rotation to obtain the solution for our hamiltonian $H_{\mathrm{eff}}$,
the wave function is
\[
\langle\rho,\phi|\gamma_{0}\rangle=[\tilde{\xi}_{1}(\rho,\phi),\tilde{\xi}_{2}(\rho,\phi),\tilde{\xi}_{2}^{*}(\rho,\phi),-\tilde{\xi}_{1}^{*}(\rho,\phi)]^{T},
\]
with ($u_{0}\equiv\exp[-\int_{0}^{\rho}d\rho^{\thinspace\prime}\Delta(\rho^{\thinspace\prime})]$)
\begin{align*}
\tilde{\xi}_{1}(\rho,\phi) & =e^{i\pi/4}e^{im\phi}u_{0}(\rho)J_{m}(\mu_{\mathrm{eff}}\rho),\\
\tilde{\xi}_{2}(\rho,\phi) & =e^{i3\pi/4}e^{i(m+1)\phi}u_{0}(\rho)J_{m+1}(\mu_{\mathrm{eff}}\rho),
\end{align*}
and the Majorana operator for $k_{z}=0$ and $E=0$ as
\begin{eqnarray}
\hat{\gamma}_{0}=\int d^{2}\mathbf{r}\thinspace u_{0}(\rho)\!\!\left[e^{i\pi/4}e^{im\phi}J_{m}(\mu_{\mathrm{eff}}\rho)c_{\uparrow}(\rho,\phi)+e^{i3\pi/4}e^{i(m+1)\phi}J_{m+1}(\mu_{\mathrm{eff}}\rho)c_{\downarrow}(\rho,\phi)+\mathrm{h.c.}\right].
\end{eqnarray}
Note that we have the real condition for Majorana fermions in real
space $\gamma^{\dagger}=\gamma$. Here the vortex winding $n\in\mathbb{Z}$
is related to $m\in\mathbb{Z}$ by
\begin{eqnarray}
n=2m+1.
\end{eqnarray}
Note that only odd vortex windings permit zero-energy solutions at
$k_{z}=0$. Finally, we add back the $-i\partial_{z}\sigma_{z}\tau_{z}$
term. It is easy to see that the wave function for the Majorana modes
have the form
\begin{eqnarray}
\langle\rho,\phi,k_{z}|\gamma_{z}\rangle & = & \exp(ik_{z}z\sigma_{z}\tau_{z})[\tilde{\xi}_{1}(\rho,\phi),\tilde{\xi}_{2}(\rho,\phi),\tilde{\xi}_{2}^{*}(\rho,\phi),-\tilde{\xi}_{1}^{*}(\rho,\phi)]^{T}\nonumber \\
 & = & [e^{ik_{z}z}\tilde{\xi}_{1}(\rho,\phi),e^{-ik_{z}z}\tilde{\xi}_{2}(\rho,\phi),e^{-ik_{z}z}\tilde{\xi}_{2}^{*}(\rho,\phi),-e^{ik_{z}z}\tilde{\xi}_{1}^{*}(\rho,\phi)]^{T}\label{eq:kzWF}
\end{eqnarray}
and the corresponding energy $E_{\mathbf{k}}$ is given by the eigen-equation
(restore the anisotropic Fermi velocities $v_{j}$)
\begin{eqnarray}
H_{\mathrm{eff}}\gamma_{z}=E_{\mathbf{k}}\gamma_{z},\ \ E_{\mathbf{k}}=\chi_{\mathrm{M}\ell}v_{z}k_{z},
\end{eqnarray}
where the chirality of the Majorana line mode is $\chi_{\mathrm{M}\ell}=+1$.
Finally, we have the Majorana operator for $k_{z}\neq0$:
\begin{eqnarray}
\hat{\gamma}_{z}(k_{z}) & = & \int\!\!d^{2}\mathbf{r}\thinspace\gamma_{z}(\rho,\phi,k_{z})\hat{f}(\rho,\phi,k_{z})\nonumber \\
 & = & \int\!\!d^{2}\mathbf{r}\thinspace u_{0}(\rho)\Bigr[+e^{ik_{z}z}\left(e^{i\pi/4}e^{im\phi}J_{m}(\mu_{\mathrm{eff}}\rho)c_{\uparrow}(\rho,\phi,z)+\mathrm{h.c.}\right)\nonumber \\
 &  & +e^{-ik_{z}z}\left(e^{i3\pi/4}e^{i(m+1)\phi}J_{m+1}(\mu_{\mathrm{eff}}\rho)c_{\downarrow}(\rho,\phi,z)+\mathrm{h.c.}\right)\Bigr],
\end{eqnarray}
which satisfies $\gamma_{z}(k_{z})=\gamma_{z}^{\dagger}(-k_{z})$
for (real) Majorana fermion.

We note that in the above derivative the condition $n=2m+1$ is needed
only for the existence of Majorana zero mode at $k_{z}=0$. This condition
is however not necessary for the existence of Majorana chiral modes
along the vortex line. With the chiral properties of Majorana modes
the above solution can be generalized to the case with a generic winding
number $n={\cal N}$. Actually, such a vortex line is topologically
equivalent to $|{\cal N}|$ vortex lines with unity winding $n={\rm sgn}({\cal N})$.
In the later case each vortex line hosts a branch of chiral Majorana
modes. Due to the chiral property putting the $n$ branches of vortex
modes together cannot annihilate them, yielding $n$ chiral Majorana
vortex modes.

The chirality of the Majorana vortex modes depends on a couple basic
properties. First, we consider a Weyl node with opposite chirality
$\chi$ without sign change in $\mu_{\mathrm{eff}}$, which is associated
with the case of an inversion symmetric Weyl metal for $\varphi_{0}=0$.
Here the transformation is given by that of the particle-hole symmetry
$\mathcal{C}$ is the operator for particle-hole symmetry such that
$\mathcal{C}H_{\mathrm{eff}}(\mathbf{k})\mathcal{C}^{-1}=H_{\mathrm{eff}}(-\mathbf{k})$.
Then we can readily show that a superconducting Weyl node of opposite
chirality {[}the hamiltonian is given by $H_{\mathrm{eff}}(-\mathbf{k})${]}
has eigenvalues $E=-v_{z}k_{z}$ for the corresponding eigenvectors
$\mathcal{C}\gamma_{z}$, or simply $H_{\mathrm{eff}}(-\mathbf{k})\mathcal{C}\gamma_{z}=(-k_{z})\mathcal{C}\gamma_{z}.$
(ii) Next we consider a Weyl node with opposite chirality $\chi$
with sign change in $\mu_{\mathrm{eff}}$, which is associated with
the case of an inversion symmetry broken Weyl metal for $\varphi_{0}\neq0$
and $\mu=0$. The corresponding transformation is $\mathcal{P}=\tau_{z}$
such that $\mathcal{P}H_{\mathrm{eff}}(\mathbf{k},\mu_{\mathrm{eff}})\mathcal{P}^{-1}=-H_{\mathrm{eff}}(-\mathbf{k},-\mu_{\mathrm{eff}})$.
(iii) Finally, we consider flipping the vortex winding $n\rightarrow-n$.
The corresponding transformation is $U_{1}=i\tau_{y}$ such that $U_{1}H_{\mathrm{eff}}(\mathbf{k},n=1)U_{1}^{-1}=H_{\mathrm{eff}}(\mathbf{k},n=-1)$.
Collecting all these results together, and restoring the anisotropic
Fermi velocities $\mathbf{v}$, that is $H_{\mathrm{eff}}(\mathbf{k})\rightarrow H_{\mathrm{eff}}(\mathbf{v}\cdot\mathbf{k})$,
we thus have
\begin{equation}
\chi_{M\ell}=\mathrm{sgn}(nv_{x}v_{y}v_{z}).
\end{equation}

\section*{III. Chiral Majorana vortex modes from real-space calculation}

\begin{figure}[h]
\begin{centering}
\includegraphics[width=0.6\textwidth]{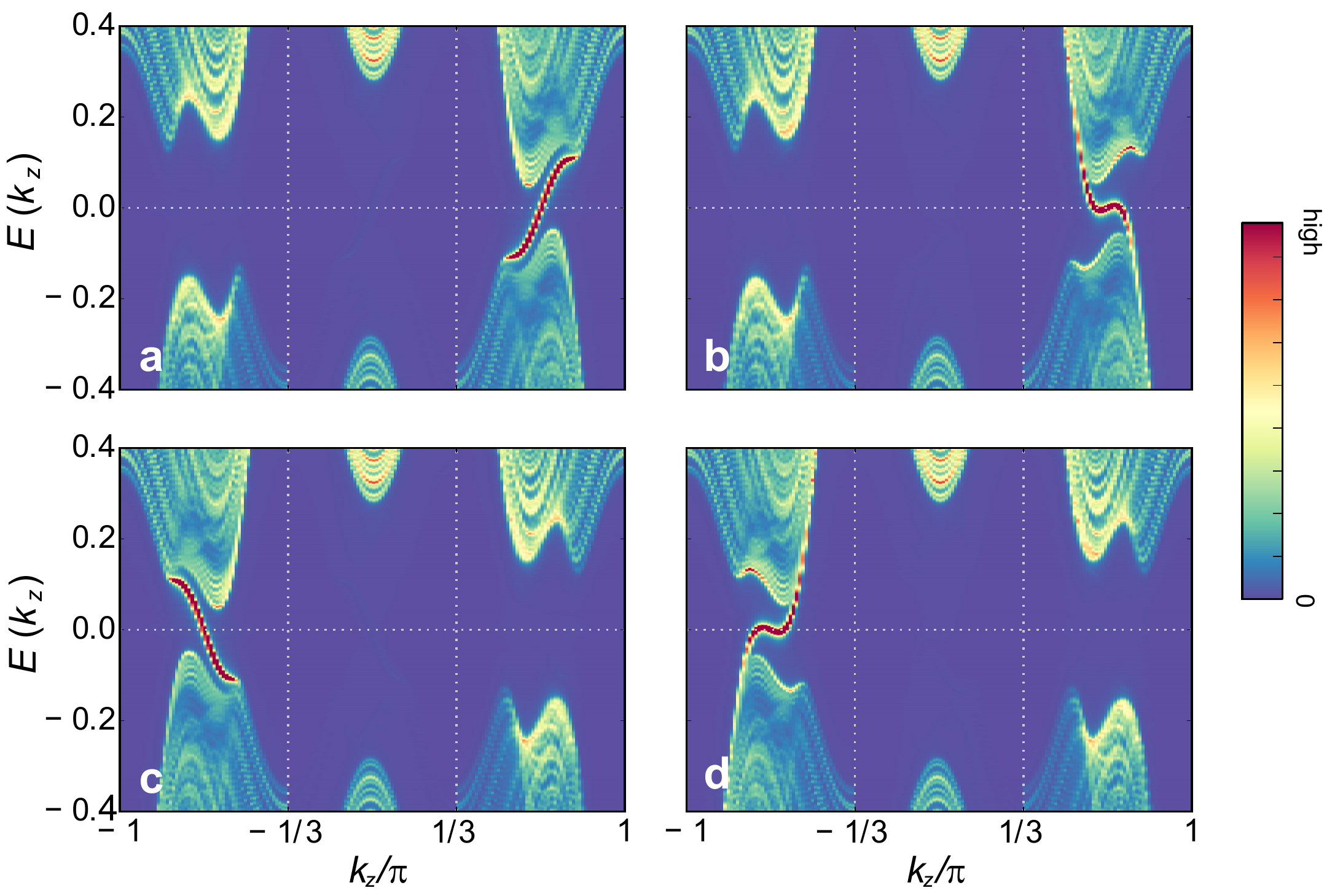}
\par\end{centering}
\caption{\label{fig:xMLM-z}Local spectral densities $A(k_{z},E)$ and chiral
Majorana line modes in inversion symmetric Weyl metal. (a) The vortex
line with positive winding number $n=+1$ and attached to $\Delta_{+2Q}$;
(b) The vortex line with negative winding number $n=-1$ and attached
to $\Delta_{+2Q}$; (c) The vortex line with positive winding number
$n=+1$ and attached to $\Delta_{-2Q}$; (d) The vortex line with negative
winding number $n=-1$ and attached to $\Delta_{-2Q}$. The parameters
are that $Q=2\pi/3$, $\mu=0.7$, $U=5.8$, with which the self consistent
mean field solutions give $\Delta_{0}=-1.849\times10^{-2}$ and $\Delta_{\pm2Q}=0.2014$.
System is periodic in $x$- and $y$-axis, while open in $z$-axis,
and has size $N_{x}/2=N_{y}=55$ and $N_{z}=162$.}
\end{figure}

\begin{figure}[h]
\begin{centering}
\includegraphics[width=0.6\textwidth]{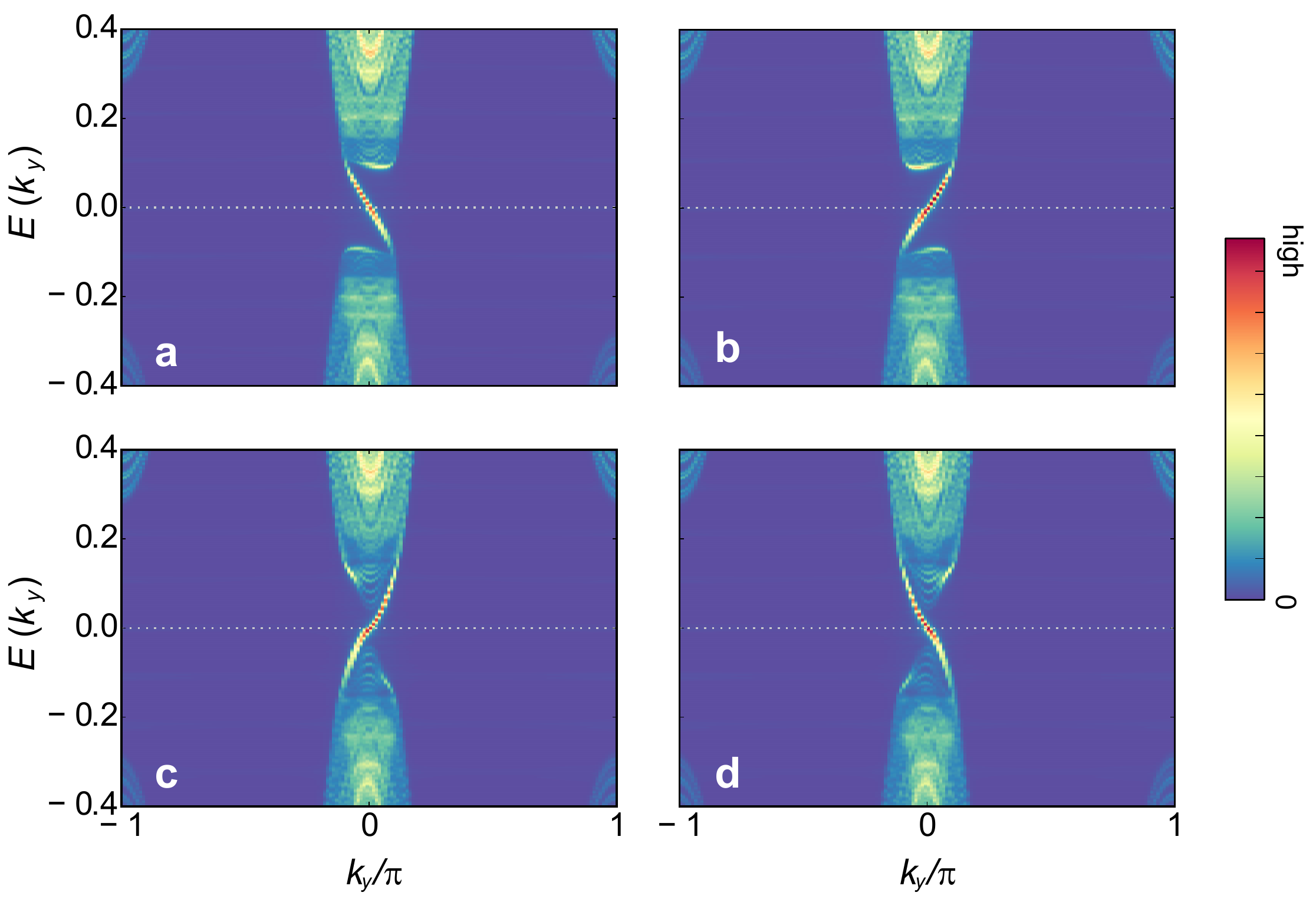}
\par\end{centering}
\caption{\label{fig:xMLM-y}Numerical results for the configurations similar
to those in Fig.\,\ref{fig:xMLM-z}, except that the vortex lines
are along $y$-axis. The system is also periodic in $x$- and $z$-axis,
while open in $y$-axis, and has size $N_{x}/2=N_{z}=54$ and $N_{y}=162$.
Besides the chiral Majorana modes, the non-chiral Andreev bound modes
are also explicitly obtained in the vortex lines.}
\end{figure}

For a full real-space diagonalization of the vortex line modes, we
consider a vortex line ($n=1$) and anti-vortex line ($n=-1$) along
$z$-axis, separated from each other in $x$-$y$ plane, and attached
to one of $\Delta_{0,\pm2Q}$. Note that in considering a vortex and
anti-vortex line pair for the system, we can still apply the periodic
boundary condition for the numerical calculation. Nevertheless, the
momentum $\mathbf{k}$ is no longer good quantum number, and so the
Hamiltonian $H_{{\rm MF}}$ should be diagonalized in real space $\mathbf{r}=(x,y,z)$.
The retarded Green's function can be obtained by
\begin{eqnarray}
\hat{G}^{R}(E)=|\alpha_{1}\rangle[\hat{G}^{R}(E)]_{\alpha_{1}\alpha_{2}}\langle\alpha_{2}|,
\end{eqnarray}
with basis in the position and spin space $\alpha\equiv(\mathbf{r},s)$
and $s=\{\uparrow,\downarrow\}$. The elements read
\begin{eqnarray}
[\hat{G}^{R}(E)]_{\alpha_{1}\alpha_{2}}=\sum_{\eta}\frac{U_{\alpha_{1}\eta}U_{\alpha_{2}\eta}^{*}}{E-E_{\eta}+i0^{+}}.
\end{eqnarray}
Here $E_{\eta}$ and $U_{\alpha\eta}$ are the eigen-energies and
corresponding eigenvector matrix obtained after solving the real space
Bogoliubov-de Gennes equation for \eqref{eq:H_MF}. Then the spin-dependent
local spectral functions are given by
\begin{eqnarray}
A_{s}(x,y,k_{z},E)=-\frac{1}{\pi}\mathrm{Im}\langle x,y,k_{z},s|\hat{G}^{R}(E)|x,y,k_{z},s\rangle,
\end{eqnarray}
where $|x,y,k_{z},s\rangle=\frac{1}{\sqrt{N_{z}}}\sum_{z}e^{ik_{z}z}|\mathbf{r},s\rangle$.
We emphasize that due to the PDW order, the momentum $k_{z}$ of the
unfolded BZ is actually not a quantum number either. Nevertheless,
to reveal the dispersion of the Majorana modes in the vortex line,
we still extract the information of relation between the energy of
vortex modes and quantum momentum $k_{z}$. Finally, the local spectral
function is
\begin{eqnarray}
A(x,y,k_{z},E)=\sum_{s}A_{s}(x,y,k_{z},E).
\end{eqnarray}
Computing $A(k_{z},E)$ near the vortex core gives the energy spectra
of the bulk and the corresponding vortex lines.

Fig.~\ref{fig:xMLM-z}(a-d) show the spectra for different configurations.
It is clear that in both cases the chiral Majorana vortex modes traverse
the bulk gap connecting the lower and upper bands. The chirality of
these modes depends on the vortex line winding number and also which
pairing order the vortex line is attached to, consistent with the
previous analytic solution. As a comparison, we have also done a similar
measurement $A(k_{y},E)$ for the vortex lines along $y$-axis and
the sign flip property of $\chi_{\mathrm{M}\ell}$ remains intact.
The results are shown in Fig.\,\ref{fig:xMLM-y}. These results show
that the chiral Majorana line modes observed in the Weyl system is
irrespective of the directions.

\section*{IV. Emergence of the second Chern number}

The second Chern number is defined in the 4D space, while the present
physical system is a 3D system. To have a second Chern number, we
parameterize the Hamiltonian by taking into account a phase factor
of the SC order as $\Delta_{\mathbf{q}}e^{in\theta}$, where $\theta\in[0,2\pi)$
forms a 1D periodic parameter space $S^{1}$. Together with the 3D
lattice, we construct a 4D synthetic space $T^{4}=T^{3}\times S^{1}$
spanned by $\mathsf{p}=(\mathbf{p},p_{\theta})$ with $\mathbf{p}=(p_{x},p_{y},p_{z})$
and $p_{\theta}=\theta$. Let the eigenvectors after solving the mean
field Hamiltonian $H_{{\rm MF}}(\mathsf{p})$ are $|\alpha,\mathsf{p}\rangle$
($\alpha$ as band index), we can then compute $C_{2}$ by
\begin{eqnarray}
C_{2}=\frac{1}{32\pi^{2}}\int_{T^{4}}d^{4}\mathsf{p}\thinspace\epsilon_{ijk\ell}\mathrm{Tr}[\mathtt{F}_{ij}\mathtt{F}_{k\ell}],\label{2ndChern}
\end{eqnarray}
where $\epsilon$ is the antisymmetric tensor and the gauge field
strengths $\mathtt{F}_{ij}$ are calculated based on the non-Abelian
gauge potentials $\mathtt{a}_{j}$ by
\begin{align*}
\mathtt{a}_{i}^{\alpha\beta}(\mathsf{p}) & =-i\langle\alpha,\mathsf{p}|\partial_{i}|\beta,\mathsf{p}\rangle,\\
\mathtt{F}_{ij}^{\alpha\beta}(\mathsf{p}) & =\partial_{i}\mathtt{a}_{j}^{\alpha\beta}-\partial_{j}\mathtt{a}_{i}^{\alpha\beta}+i[\mathtt{a}_{i},\mathtt{a}_{j}]^{\alpha\beta}.\\
 & =-i\left(\partial_{i}\langle\alpha,\mathsf{p}|P_{E}(\mathsf{p})\partial_{j}|\beta,\mathsf{p}\rangle-(i\leftrightarrow j)\right).
\end{align*}

Since the Chern numbers are topological invariants, they are not changed
as long as the bulk gap of the system is not closed. In this way,
the calculation of the Chern numbers can be simplified by deforming
the original bulk Hamiltonian so that the bulk bands become flat while
keeping the gap to be open. In other words, we consider the diagonalized
Hamiltonian in the flat-band limit, given by
\[
h_{F}(\mathsf{p})=\varepsilon_{G}P_{G}(\mathsf{p})+\varepsilon_{E}P_{E}(\mathsf{p}),
\]
where $\varepsilon_{G}<0$, $\varepsilon_{E}>0$ and $P_{G(E)}$ is
a projection operator to the occupied (empty) states. With the projection
operator the second Chern number can be further calculated by \cite{Bernevig2013}
\begin{equation}
C_{2}=\frac{1}{8\pi^{2}}\int_{T^{4}}d^{4}\mathsf{p}\thinspace\epsilon_{ijk\ell}\mathrm{Tr}[P_{E}\partial_{i}P_{G}\partial_{j}P_{G}P_{E}\partial_{k}P_{G}\partial_{\ell}P_{G}],\label{eq:C2}
\end{equation}
where $\partial_{i}\equiv\frac{\partial}{\partial\mathsf{p}_{i}}$.
We note that this expression is gauge invariant and it is suitable
for numerical calculation. Numerically, the derivatives are taken
using the symmetric difference quotient method, namely
\[
\partial_{j}P_{G,E}(\mathsf{p}_{i})=\frac{1}{2\delta_{j}}[P_{G,E}(\mathsf{p}_{i}+\delta_{j})-P_{G,E}(\mathsf{p}_{i}-\delta_{j})],
\]
where individual $\mathsf{p}_{i}=2\pi\frac{n_{i}}{N_{i}}$ for $n_{i}=0,1,\dots,N_{i}-1$,
and $\delta_{j}=\frac{1}{r'}\frac{2\pi}{N_{j}}$. Here $N_{i}$ is
the discretization in the $i$-th direction and $r'$ is a parameter
added to enhance the accuracy of the derivatives.

\begin{table}[h]
\begin{centering}
\begin{tabular}{|c|c|c|}
\hline
$U$  & $C_{2}$  & Note\tabularnewline
\hline
\hline
$5.6$  & $-0.89783698$  & gapless spectrum\tabularnewline
\hline
$5.7$  & $-0.9554145$  & Fully gapped\tabularnewline
\hline
$5.8$  & $-0.95628694$  & Fully gapped\tabularnewline
\hline
$5.9$  & $-0.95659966$  & Fully gapped\tabularnewline
\hline
$6$  & $-0.9561468$  & Fully gapped\tabularnewline
\hline
\end{tabular}
\par\end{centering}
\caption{\label{tab:C2_U}Second Chern numbers for positive vortex winding
$n=+1$ attached to $\Delta_{+2Q}$. We consider inversion symmetric
Weyl metal with $\mu=0.7$ and various $U$'s. The system size is
$N_{x,y,z,\theta}=N_0=100$ and derivative enhancing parameter is $r'=5$.}
\end{table}

\begin{table}[h]
\begin{centering}
\begin{tabular}{|c|c|c|}
\hline
Vortex lines attached to  & Vortex line winding $n$  & $C_{2}$\tabularnewline
\hline
\hline
$\Delta_{+2Q}$  & $+1$  & $-0.95628694$\tabularnewline
\hline
$\Delta_{+2Q}$  & $-1$  & $+0.95628694$\tabularnewline
\hline
$\Delta_{-2Q}$  & $+1$  & $+0.95628694$\tabularnewline
\hline
$\Delta_{-2Q}$  & $-1$  & $-0.95628694$\tabularnewline
\hline
\end{tabular}
\par\end{centering}
\caption{\label{tab:Sign-flip-property-C2}The 2nd Chern number $C_{2}$ with
respect to $\chi$ and $n$. We consider inversion symmetric Weyl
metal with $\mu=0.7$ and $U=5.8$. The system size is $N_{x,y,z,\theta}=N_0=100$
and derivative enhancing parameter is $r'=5$.}
\end{table}

On the other hand, if we compute the 2nd Chern number~\eqref{2ndChern}
for only a single superconducting Weyl cone with vortex line based
on the low-energy effective Hamiltonian $H_{{\rm eff}}$, as given
in~\eqref{eq:H_eff}, we can show that the corresponding second Chern
number takes a simple form
\begin{equation}
C_{2}=-\chi n\in\mathbb{Z}.\label{eq:chi_n}
\end{equation}
This result is consistent with the 2nd Chern number computed numerically
with formula~\eqref{eq:C2} and based on the full lattice Hamiltonian,
as summarized in Table \ref{tab:C2_U}. We also find that when the
system size increases, the magnitude of $|C_{2}|$ approaches unity.
The sign flip property of $C_{2}$ is shown in Table \ref{tab:Sign-flip-property-C2}.

\section*{V. Discussions}

We note that in defining the synthetic 4D space, the parameter $p_\theta$ for the SC order is a constant variable, independent of real position $\bold r$. This definition has nothing to do with the vortex line generated in the real space. On the other hand, once the superconducting Hamiltonian $H_{{\rm MF}}(\mathbf{p},p_{\theta})$ for the defined 4D synthetic space is topologically nontrivial, namely, has nonzero 2nd Chern number, it gives rise to novel physical consequences, including that a vortex line generated in the PDW phase hosts chiral Majorana modes.

The emergence of the 2nd Chern number shows that the gapped PDW phase, while being trivial
in the 3D, is an intrinsic non-Abelian topological order in the 4D synthetic
space. Nevertheless, we note that this emergent topological phase is beyond the ten-fold Altland-Zirnbauer symmetry
classification~\cite{Altland-Zirnbauer}. Actually, within the scope of the current topological classification theory~\cite{PhysRevB.78.195125,kitaev2009,Teo2016RevModPhys}, the class-D superconductor in 4D should still be topologically trivial, implying that the 4D Hamiltonian $H_{{\rm MF}}(\mathbf{p},p_{\theta})$ cannot be characterized by the current topological classification theory. A subtle reason is because the extra dimension, defined through the parameter space $p_\theta$ of the SC order, is qualitatively different from the 3D physical space $(p_x,p_y,p_z)$. This leaves an interesting open question: how to classify the topological states in generic synthetic dimensions.

\appendix

\bibliographystyle{apsrev4-1}
\bibliography{refs}

\end{document}